\documentclass[%
 reprint,
superscriptaddress,
frontmatterverbose, 
showpacs,preprintnumbers,
 amsmath,amssymb,
 aps,
pra,
]{revtex4-1}
\usepackage{subfigure}
\usepackage{xcolor}
\usepackage{graphicx}
\usepackage{dcolumn}
\usepackage{bm}
\usepackage{braket}

\begin{document}

\title{Coherent Control of Fano Resonances in a Macroscopic Four-Mirror Cavity}

\author{Nikhil Pramanik }
\affiliation{School of Physics, University of Hyderabad, Hyderabad, Telangana 500046, India} 
\author{K. C. Yellapragada}
 \affiliation{Centre for Quantum Technologies, 3 Science Drive 2, National University of Singapore, 117543, Singapore}
 \author{Suneel Singh}
 \affiliation{School of Physics, University of Hyderabad, Hyderabad, Telangana 500046, India} \author{P. Anantha Lakshmi}
 \email{palsp@uohyd.ernet.in}
 \affiliation{School of Physics, University of Hyderabad, Hyderabad, Telangana 500046, India}%
\date{\today}

\begin{abstract}

We demonstrate coherent control of optomechanically induced transparency and Fano resonances in a four mirror macroscopic optomechanical cavity, with two movable mirrors, each driven by an external mechanical pump.  The variable control of the amplitude and phase of the coherent mechanical pumps  provides a means of tuning the shape and nature of the  Fano profiles.  Further, our scheme shows the occurrence of tunable optomechanical features, even at very low mechanical driving field amplitudes, in macroscopic optomechanical cavities.

\begin{description}

\item[Keywords]
 Four mirror, Fano resonance, optomechanically induced transparency, mechanical driving, optomechanical system   
\end{description}
\end{abstract}
\maketitle

\section{Introduction} 
Quantum interference between different transition pathways gives rise to several interesting physical phenomena, such as Fano resonances \cite{fano1961,ravirau2004}, that have been observed in a variety of physical systems. For instance, in plasmonics, photons are allowed to travel through multiple transition pathways which interfere, thus making the occurrence of Fano line shapes quite common in such materials.  Fano resonances have been observed in a wide variety of systems which include the phonon interactions in solids \cite{scott1974,hase2006}, electron transport in quantum wells, quantum dots \cite{faist1997,kroner2008}, 3D waveguides \cite{gunupudi2019},   coupled photonic microcavities \cite{smith2004,fan2002}, plasmonic metamaterials \cite{luk2010,liu2009} and nanostructures \cite{yoon2012,sasaki2009,Johnson2004,kobayasi2004} and photonic materials \cite{rahmani2012,rybin2009,ding2012,nojima2014,shang2014}.

Fano line shapes have  also been used in obtaining information on the interaction between a wide variety of nanostructures with light \cite{gallinet2011}, for local refractive index sensing applications \cite{verellen2009}, efficient confinement of light \cite{Miroshnichenko2010}, surface enhanced Raman scattering (SERS) \cite{ye2012}, generation of slow light  in metamaterials \cite{Wu2011}, enhanced light transmission \cite{zhou2011} and sensitive biosensors \cite{Lee2011}.  Fano interference is seen to play an important role in  producing lasing without population inversion \cite{arkhipkin1983,kocharovskaya1988,scully1989,fry1993} and as a tool to probe decoherence \cite{agarwal1984,barnthaler2010} in the  field of quantum optics and quantum information.  More recently, Fano resonances have been widely studied theoretically in hybrid optomechanical systems with distinct configurations involving double cavities \cite{qu2013}, whispering gallery modes \cite{zhang2017}, BEC \cite{akram2017,Yasira2016}, two level atom/qubit \cite{akram2015,jiang2017}, to name a few.     A detailed analysis of Fano resonances and the generation of slow light  was carried out in Ref.\cite{akram2017,jiang2017}.
 
In the existing studies of cavity optomechanics, the optomechanical (OM) effects (that depend upon $G^2$) become significant only when the effective OM coupling parameter $G$ is sufficiently large.  This in turn requires extremely small sizes of mirrors and cavity arm lengths because the OM coupling parameter is inversely proportional to cavity length, size and mass (physical dimensions) of the mirror.  Obviously in macroscopic cavities i.e. for large cavity length, mirror size and weight, OM coupling $G$ is very weak and hence it is not possible to observe OM effects for such small values of $G$.

In this work we propose a novel scheme that enables one to observe OM effects even in macroscopic cavities,  i.e., even for negligible G.  We study the occurrence of Fano resonances and the related phenomenon of optomechanically induced transparency (OMIT) \cite{weissOMIT} by analyzing the fields generated at the anti-Stokes frequencies in a macroscopic cavity.  By introducing  coherent mechanical pump to act on two mirrors of the four mirror cavity, we show that  optomechanical interaction can be enhanced, thus resulting in appearance of Fano resonances and OMIT features in the generated anti-Stokes signals.  For suitable choice of amplitude and phase of the mechanical driving fields and mirror oscillation frequencies, we demonstrate the occurrence of tunable double Fano resonance  in a  macroscopic four mirror OM cavity.  From a detailed study, we further identify the interfering contributions to the fields generated at anti-Stokes frequencies.
 
The paper is organized as follows: In Sec. II a detailed description of 
the system that is studied here is provided, together with necessary mathematical formulation and solutions for the  dynamical evolution of different quantities of interest.   In Sec. III numerical results pertaining to the OMIT and the asymmetric Fano-like resonance induced by interference of two transition pathways are presented.   We further show the generation of double Fano-like resonances  for specific choice of the drive amplitudes and phases.   A summary of results and conclusions are presented in Sec. IV.

\section{\label{sec:level2}Model and Theory }

\begin{figure}[ht]
 	\includegraphics[width=\linewidth,keepaspectratio]{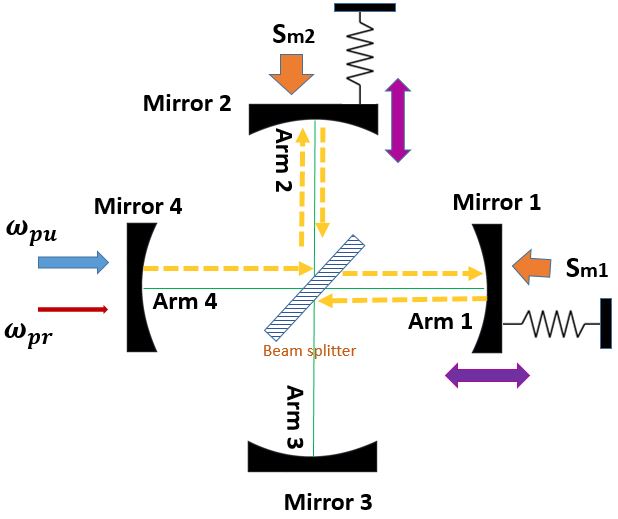}
	\caption{Schematic diagram of a four mirror cavity with two movable mirror.}
	\label{fig:1}
\end{figure}

Fig. \ref{fig:1} shows  a schematic of the four mirror cavity considered in this work.
Mirrors 1 and 2 are movable, each of which are driven by a coherent mechanical pump while mirrors 3 and 4 are fixed.  A pump laser of frequency $\omega_{pu}$  and a probe laser of frequency $\omega_{pr}$  enter the cavity from the left.  The frequency of the cavity is taken to be $\omega_0$.  Depending on which cavity is experiencing stronger field (as decided by the reflectivity and transmittivity of the beamsplitter and the optical length of each of the arms) will correspondingly exhibit stronger optomechanical coupling resulting from the radiation pressure, produced due to the incident input field inside the cavity.   In this study, the parameters are chosen in  such a manner as to render the field in arm 3 negligible \cite{farman2014}.

The Hamiltonian, describing various interactions that are considered here, is written in a rotating frame with frequency $\omega_0$ in Eq. (\ref{eq:H}). 

\begin{equation}
\begin{aligned}
H  = ~  &\hbar\Delta_c a^{\dagger}a+\sum_{i=1,2} \left( \frac{p_i^2 }{2m_i} + \frac{1}{2}m_i\omega_{mi}^2x_i^2 -\hbar g a^\dagger a x_i\right) \\
& + i\hbar\epsilon_{pu}(a^\dagger - a) + i\hbar \epsilon_{pr} (a^\dagger e^{-i\delta t}-a e^{i\delta t}) \\
&-\sum_{i=1,2}s_{mi}x_i cos(\delta t + \phi_{mi})
\end{aligned}
\label{eq:H}
\end{equation}

Here $a$ and $a^\dagger$ are the bosonic operators of the cavity field,  $p_i$ and $x_i$ are the momentum and position variables of the two movable mirrors each of which are modeled as simple harmonic oscillators, with frequency $\omega_{mi}$, effective
mass $m_i$ and mechanical decay rate $\gamma_{i}$  with $i$ taking the values 1,2. The pump (probe) amplitude  $\epsilon_{pu}$ ($\epsilon_{pr}$) is related to the input pump (probe) power $P_{pu}$ ($P_{pr}$) as
$\epsilon_{pu}=\sqrt{2\kappa P_{pu}/\hbar \omega_{pu}}$  ($\epsilon_{pr}=\sqrt{2\kappa P_{pr}/\hbar \omega_{pr}}$). The optomechanical coupling constant $g$  is given by $g = {\omega_{pu}}/({L_4+L_1t^2+L_2r^2})$ where $L_i$  ($i = 1,4$)  are the respective arm lengths of the  four mirror setup with  $r$ and $t$ the reflection and transmission coefficients of the beam splitter.  The quantities $s_{mi}$ and $\phi_{mi}$ $(i=1,2)$ are the amplitude and phase of the coherent mechanical drive. 

It is often convenient to work with dimensionless position and momentum operators. We now define the dimensionless position and momentum operators for each of the two mirrors (Eq. (\ref{eq:dimless})) and rewrite the Hamiltonian in terms of these operators as given by Eq. (\ref{eq:H2}).

\begin{equation}
\begin{aligned}
x_i=\sqrt{\frac{\hbar}{m_i \omega_{mi}}} \; Q_i \;\: \text{and} \;\: p_i=\sqrt{m_i \hbar \omega_{mi}} \; P_i
\end{aligned}
\label{eq:dimless}
\end{equation}

\begin{equation}
\begin{aligned}
H  = \;  &\hbar \;\Delta_c \; a^{\dagger}a+ \sum_{i=1,2}\frac{\hbar  \omega_{mi}}{2} \; ( \; P_i^2+Q_i^2 \; )  -\hbar\sum_{i=1,2} G_{i} \; a^\dagger a Q_i\\
& + i\hbar \;\epsilon_{pu} \; (a^\dagger - a)+ i\hbar \; \epsilon_{pr} \; (a^\dagger \; e^{-i\delta t}-a \; e^{i\delta t})  \\
& - \sum_{i=1,2}S'_{mi} ~  Q_i \; cos\; ( \; \delta t + \phi_{mi} \;)
\end{aligned}
\label{eq:H2}
\end{equation}

Here the first term describes the cavity field energy with  the cavity detuning given by $\Delta_c=\omega_0 - \omega_{pu}$ and the second term is the energy of the two mechanical oscillators.  The third term describes  the optomechanical interaction arising due to the coupling between the two mechanical oscillators and the cavity field. The effective coupling coefficient between cavity field and mirror 1 and 2 respectively are given by $G_1=t^2g \sqrt{\hbar/m_1 \omega_{m1}}$ and $G_2=r^2g\sqrt{\hbar/m_2 \omega_{m2}}$. The fourth and fifth terms describe the interaction between the cavity field and the input pump and probe fields respectively,  with $\delta=\omega_{pr}-\omega_{pu}$, the pump - probe detuning.  The last term describes the mechanical pumping energy applied to  each movable mirror, with  the driving parameter $S'_{mi}$  defined by $S'_{mi}=s_{mi}\sqrt{\hbar/m_i \omega_{mi}}$ (i=1,2).
 
 We denote the expectation values of each of the operators $\hat a$, $\hat Q$ and  $\hat P$ with $\braket{a}$, $\braket{Q}$ and $\braket{P}$ respectively.
  Using the Hamiltonian given in Eq. (\ref{eq:H2}), we derive the equations that describe the dynamical behaviour of these operator expectation values in Heisenberg picture, as follows. 
 \begin{equation}
 \begin{aligned}
 \frac{d\braket{a}}{dt} = &-i\Delta_c \braket{a} + \epsilon_{pu} + \epsilon_{pr} e^{-i\delta t} + iG_1 \braket{a} \braket{Q_1}\\
 & + iG_2\braket{a}\braket{Q_2}-\kappa \braket{a}\\
 \frac{d\braket{Q_i}}{dt}= & ~ \omega_{mi}\braket{P_i}, (i=1,2)\\
 \frac{d\braket{P_i}}{dt}=&-\omega_{m1}\braket{Q_i} + G_i \braket{a^\dagger} \braket{a} - \gamma_i \braket{P_i}\\
 & + \frac{S'_{mi}}{\hbar}cos(\delta t+\phi_{mi}), (i=1,2). \\
 \end{aligned}
 \label{eq:HL}
 \end{equation}
 In the above equations,  $\kappa$ is the cavity decay rate.  In obtaining the above equations of motion,  
 the mean field assumption $\braket{MN}=\braket{M}\braket{N}$  for the relevant operators has been used.  Next, each of the variables is  separated into a steady state solution and a small fluctuation around its steady state value,  i.e., $\hat{x}=x_s+\delta \hat{x}$.  
 By substituting the same in each of the above equations,  we can obtain the steady state solutions (Eq. (\ref{eq:SS})) as well as the equations of motion for the  fluctuations in  each of these operators.  The steady state solutions are given by 
\begin{equation}
\begin{aligned}
&P_{is}=0, (i=1,2)\\
&a_s= \frac{\epsilon_{pu}}{\kappa + i\Delta}\\
&Q_{is}=\frac{G_i\left|a_s\right|^2}{\omega_{mi}}, (i=1,2) \\
\end{aligned}
\label{eq:SS}
\end{equation}
where $\Delta=\Delta_c-G_1Q_1-G_2Q_2$. The equations of motion for the fluctuations in each of the operators are obtained as 

  \begin{widetext}
	\begin{eqnarray}
   (\frac{d}{dt}+(\kappa +i \Delta_c ))\delta a & = & \epsilon_{pr}e^{-i\delta t} + i G_1 (Q_{1s}\delta a+a_s\delta Q_1)
 +iG_2(a_s\delta Q_2+ Q_{2s}\delta a)\\
  (\frac{d^2}{dt^2}+\gamma_1 \frac{d}{dt}+\omega_{m1}^2)\delta Q_1& = & G_1\omega_{m1}(a_s^*\delta a+a_s\delta a^*)
 +\frac{S'_{m1}\omega_{m1}}{\hbar}cos(\delta t+\phi_{m1})\\
  (\frac{d^2}{dt^2}+\gamma_2 \frac{d}{dt}+\omega_{m2}^2)\delta Q_2& = & G_2\omega_{m2}(a_s^*\delta a+a_s\delta a^*)
 +\frac{S'_{m2}\omega_{m2}}{\hbar}cos(\delta t+\phi_{m2}).
 \end{eqnarray}
 \end{widetext}
  
  We next  use the ansatz
  \begin{equation}
  \begin{aligned}
  &\delta X_i=X_i^- e^{-i\delta t} + X_i^+ e^{i\delta t}, (i=1,2), \\
  \end{aligned}
  \label{eq:ansatz}
  \end{equation}
   
   for each of the variables and substitute this into the equations of motion for the fluctuations and group the coefficients of like-terms to obtain the solutions for the relevant quantities of interest.   For example, the term $a_1^-$ is of interest in obtaining the  anti-Stokes component of the output field from the cavity.   In particular,  the real part of anti-Stokes field is given by  $Re(\eta_{as})=Re(2\kappa a_1^-/\epsilon_{pr})$ \cite{qu2013}.  The resulting equations for the fluctuations in the relevant variables are obtained as 
   
  \begin{widetext}
  	\begin{eqnarray}
  	(\kappa +i(\Delta-\delta))a_1^- &=& \epsilon_{pr} + iG_1a_sQ_1^- + iG_2a_sQ_2^-\\
  	(\kappa +i(\Delta+\delta))a_1^+ &=& iG_1a_sQ_1^+ + iG_2a_sQ_2^+\\
  	(\omega_{m1}^2-i\gamma_1\delta-\delta^2)Q_1^- &=& G_1 \omega_{m1}(a_s^* a_1^- + a_s (a_1^+)^*)+\frac{S'_{m1}\omega_{m1}}{2\hbar}e^{-i\phi_{m1}}\\
  	(\omega_{m1}^2+i\gamma_1\delta-\delta^2)Q_1^+ &=& G_1 \omega_{m1}(a_s^* a_1^+ + a_s (a_1^-)^*)+\frac{S'_{m1}\omega_{m1}}{2\hbar}e^{+i\phi_{m1}}\\
  	(\omega_{m2}^2-i\gamma_2\delta-\delta^2)Q_2^- &=& G_2 \omega_{m2}(a_s^* a_1^- + a_s (a_1^+)^*)+\frac{S'_{m2}\omega_{m2}}{2\hbar}e^{-i\phi_{m2}}\\
  	(\omega_{m2}^2+i\gamma_2\delta-\delta^2)Q_2^+ &=& G_2 \omega_{m2}(a_s^* a_1^+ + a_s (a_1^-)^*)
  	+\frac{S'_{m2}\omega_{m2}}{2\hbar}e^{+i\phi_{m2}},
  	\end{eqnarray}
	
   the solutions of which are obtained as 
\begin{eqnarray}
Q_2^- &=& \frac{\epsilon_{pr}a_s^* G_2 \chi_2(\delta)\alpha + 2 \Delta\chi_2(\delta) G_1 G_2 \left|a_{s}\right|^2 Q_1^- + S_{m2} \chi_2(\delta)  \alpha \beta e^{-i \phi_{m2}}}{\alpha \beta - 2 G_2^2 \left|a_{s}\right|^2 \chi_2(\delta) \Delta }\\\
Q_1^- &=& \frac{\alpha\chi_1(\delta) [\epsilon_{pr}a_s^*  G_1 (d_2+2\Delta G_2^2\left|a_{2s}\right|^2\chi_2(\delta))
+ \beta S_{m1} e^{-i\phi_{m1}}d_2]
+\alpha \beta \chi_1(\delta)\chi_2(\delta)2\Delta G_1 G_2 \left|a_{2s}\right|^2 S_{m2} e^{-i\phi_{m2}}}{d_1 d_2 - 4 \Delta^2 G_1^2 G_2^2 \left|a_s\right|^4\chi_1(\delta)\chi_2(\delta)}\\
a_1^- &=& \frac{\epsilon_{pr}+iG_1 a_s Q_1^- + i G_2  a_s Q_2^-}{\kappa + i (\Delta -\delta)}
\label{eq:final2}
\end{eqnarray}


Each of the hitherto undefined quantities that appear in the above solutions are defined by

\begin{equation}{\nonumber}
\alpha = (\kappa-i(\Delta+\delta));~~~ \beta = (\kappa+i(\Delta-\delta));~~~
d_1=\alpha \beta -2 G_1^2  \left|a_s\right|^2\chi_1(\delta)\Delta;~~~d_2=\alpha \beta  -2 G_2^2  \left|a_s\right|^2\chi_2(\delta)\Delta;
\end{equation}
\begin{equation}
\Delta=\Delta_c-G_1Q_1-G_2Q_2;~~~S_{mi} = \frac{S'_{mi}}{2\hbar}, (i=1,2);~~~
\chi_1(\delta)=\frac{\omega_{m1}}{\omega_{m1}^2-i\gamma_1\delta-\delta^2} ~~~\text{and} ~~\chi_2(\delta)=\frac{\omega_{m2}}{\omega_{m2}^2-i\gamma_2\delta-\delta^2}~.
\label{eq:final3}
\end{equation}
\end{widetext}

\newpage
In the next section we present numerical simulations of the results  for the real part of the generated anti-Stokes field as a function of the normalized probe detuning $\delta_{pr}/\Omega$ where $\delta_{pr}=\omega_{pr}-\omega_{0}$, for a wide range of parameters and  discussion of the results. 

\section{Results and Discussion}

Here we focus on a study of the generated anti-stokes field ($\eta_{as}$)  in the four mirror cavity, for a wide range of values of the input parameters.

\begin{figure}[h]
	\includegraphics[width=\linewidth,keepaspectratio]{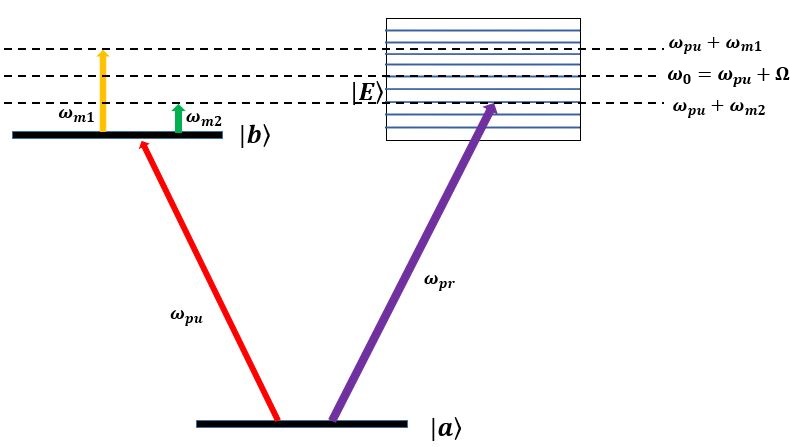}
	\caption{Schematic illustration of frequencies used
		in obtaining Fano line shapes.}
	\label{fig:grapha} 
\end{figure}

As shown in  Fig. \ref{fig:grapha}, when the pump field of frequency $\omega_{pu}$ interacts with the mechanical mirror of frequency, say, $\Omega$, absorption and emission of phonons creates the anti-Stokes field ($\omega_{pu} + \Omega$) and the Stokes field ($\omega_{pu} - \Omega$) respectively \cite{gsagarwalbook}. If the frequency of pump laser is red-detuned (tuned below the resonance frequency of cavity exactly by an amount  $\Omega$), then the anti-Stokes field becomes resonant  with the cavity field and therefore gets enhanced,  at the cost of suppression of the Stokes field, as now the Stokes field is far removed from resonance.  

Ugo Fano, in a seminal paper \cite{fano1961}, first described the asymmetric profile resulting in Rydberg spectral atomic lines and provided a detailed explanation of this feature arising due to resonant destructive interference between two transition pathways.  In particular, one observes a  minimum with an accompanying maximum very close to it,  which is known by the name  Fano resonance.  Inside the four mirror cavity,  when the frequency of the probe beam is tuned to that of the generated anti-Stokes field, destructive interference between these two fields gives rise to Fano-like resonance.  As shown in Fig. \ref{fig:grapha},  interference takes place between  the two transition pathways $\ket{a}\rightarrow\ket{E}$ and $  \ket{a}\rightarrow\ket{b}\rightarrow\ket{E}$, where $\ket{E}$ represents a continuum of states.   Thus, when $\omega_{pr}=\omega_{pu}+\omega_{m1}$, destructive interference between the two fields (anti-Stokes and the probe)  leads to a Fano-like resonance.  
Availability of  two oscillating mirrors, which can be tuned independently of each other,  at frequencies $\omega_{m1}$ and $\omega_{m2}$ respectively, provides  two transition pathways  from state $\ket{b}$  to the continuum $\ket{E}$.  A suitable choice of these parameters gives rise to a double Fano-like resonance.  For instance, choosing the mechanical oscillation frequency of  both the mirrors to be equal ($\omega_{m1}=\omega_{m2}$) results in superposition of the two resonances. Under this condition, the resulting Fano-like  profile can be modified, giving rise to novel features, by tuning the amplitude and phase of the coherent mechanical pump.  These features are illustrated in the following figures. 

In this study, we have considered a truly macroscopic 4-mirror cavity, with the following parameters.  Each of the cavity lengths $L_i=35$ mm, ($i = 1,4$), the effective pump laser detuning $\Delta = \Omega=2\pi \times 10^7$ Hz,  the cavity decay rate $\kappa=2\pi \times 10^6$ Hz, mechanical damping rate of each mirror $\gamma_{mi} = 2\pi \times 10^4$ Hz ($i=1,2$), mass of each mirror $m_i=14.5$g, ($i = 1,2$), wavelength of pump laser $\lambda=1064$ nm and pump power $P_{pu}=10$ $\mu$W respectively.   It is to be noted that the parameters used in this study have been obtained from a careful survey of existing literature on macroscopic hybrid optomechanical systems \cite{kc2018,sekatski2014,suzuki2015,ma1,ma2,ma3,ma4,ma8,ma9,ma10}.
Due to the macroscopic nature of the parameters of the four-mirror setup,  the optomechanical coupling is very weak to show any observable optomechanical effects  such as the optomechanically induced transparency, the Fano resonances which were earlier reported  in microscopic cavities \cite{qu2013,farman2014}.   This problem can however be circumvented by application of very nominal coherent driving field, resulting in observable optomechanical effects.  In the following, we present numerical results for the real part of the anti-Stokes field for a variety of parameters, the details of which are contained in each of the figures, and demonstrate the significant role played by the amplitude and phase of the coherent mechanical drive, in observing the novel features reported here.

In the analytical expressions that were obtained in the previous section, the optomechanical interaction terms appear to several orders of the coupling parameter $G$.   However, as the present OM system which has very weak  optomechanical  coupling, owing  to its  macroscopic nature as  determined by the parameters that were considered,  the quadratic and higher order terms result in negligible contribution to the features that are observed.   We therefore neglect the  higher order optomechanical interaction terms and retain only the linear terms in the interaction strength $G$.   This gives rise to  considerable simplification in the expression for $a_1^-$,  which is obtained after  neglecting all components containing second and higher order terms of effective OM coupling strength $G_i$ ($i=1,2$), as shown below.   It is to be noted that the  scheme proposed here, involving the coherent mechanical driving of the mirrors, enhances the optomechanical interaction giving rise to observable effects, even at very nominal mechanical driving amplitudes $s_{mi}$ ($ i = 1,2$).

\begin{equation}
\begin{aligned}
a_1^- = \frac{\epsilon_{pr}+i G_1 a_s S_{m1} \chi_1(\delta) e^{-i\phi_{m1}}+i G_2 a_s S_{m2} \chi_2(\delta) e^{-i\phi_{m2}}}{\kappa + i(\Delta - \delta)}.
\end{aligned}
\label{eq:redu_eq}
\end{equation}

This expression can be simplified further by assuming $G_1=G_2=G$, $\chi_1=\chi_2=\chi$  resulting in the following. 

\begin{equation}
\begin{aligned}
a_1^- = \frac{\epsilon_{pr}+ A ( S_{m1}e^{-i\phi_{m1}}+ S_{m2}e^{-i\phi_{m2}} )}{\kappa + i(\Delta - \delta)}, 
\end{aligned}
\label{eq:small_eq1}
\end{equation}

where $A$ appearing in the above equation is given by $A = i G a_s\chi(\delta)$. We now study the behaviour of output spectrum for various combinations of amplitudes and phases of the two coherent mechanical pumps.  At first, we look at the case when $\phi_{m1}=\phi_{m2}=3\pi/2$.

\begin{figure}
	\includegraphics[width=\linewidth,keepaspectratio]{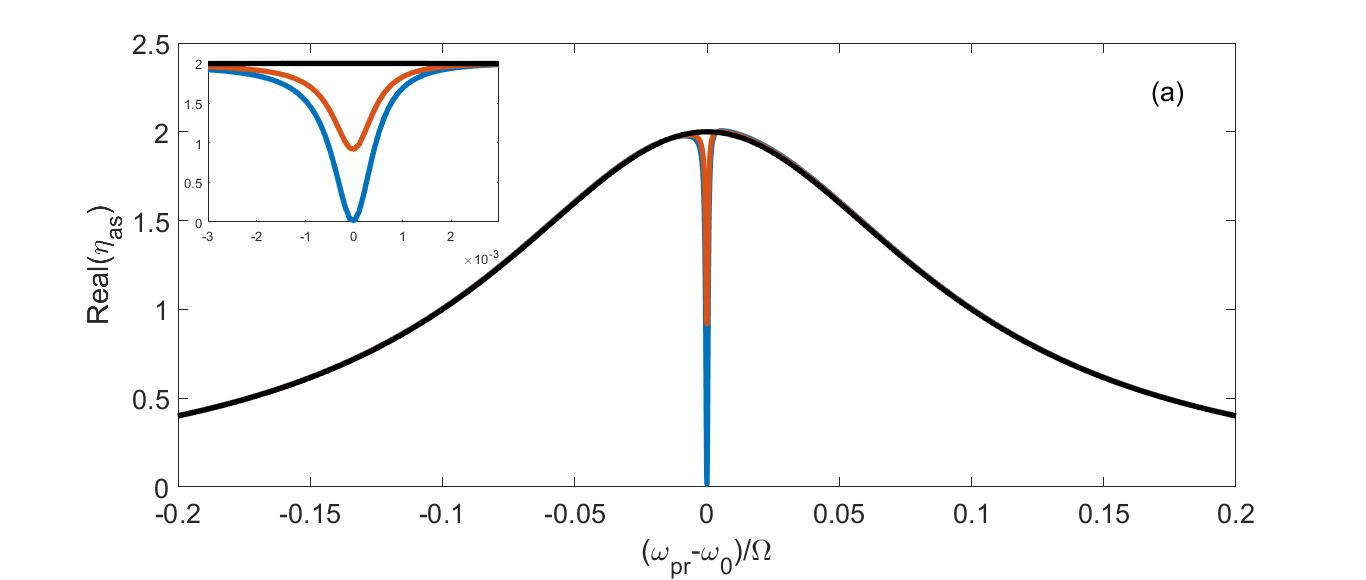}
	\includegraphics[width=\linewidth,keepaspectratio]{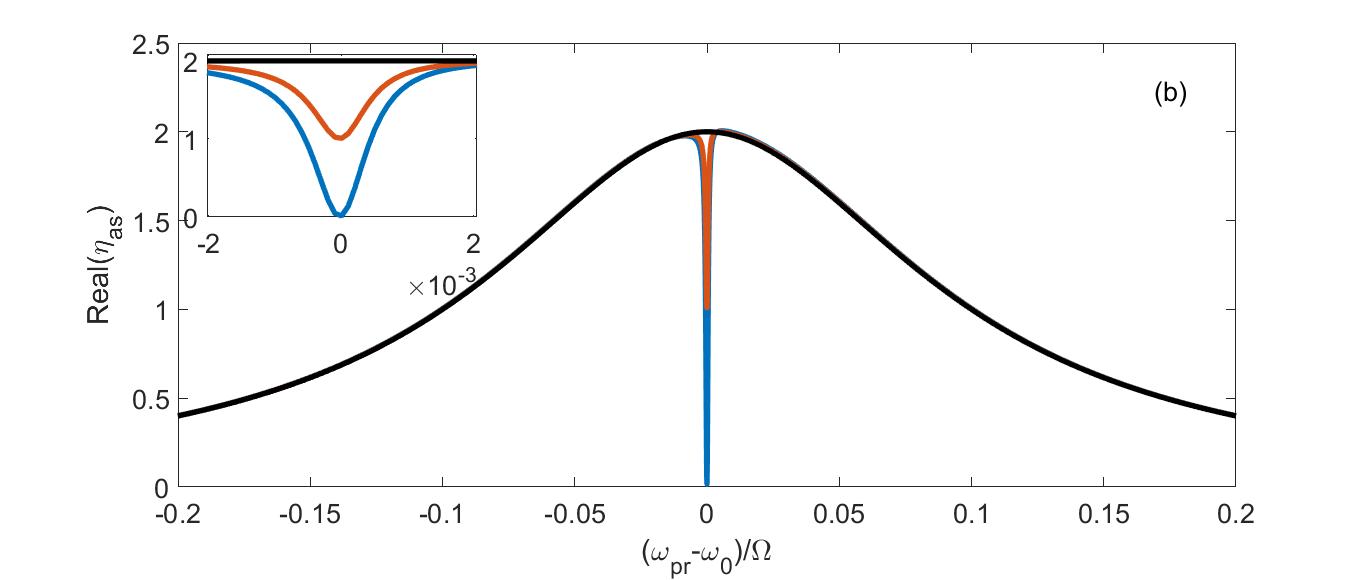}
	\includegraphics[width=\linewidth,keepaspectratio]{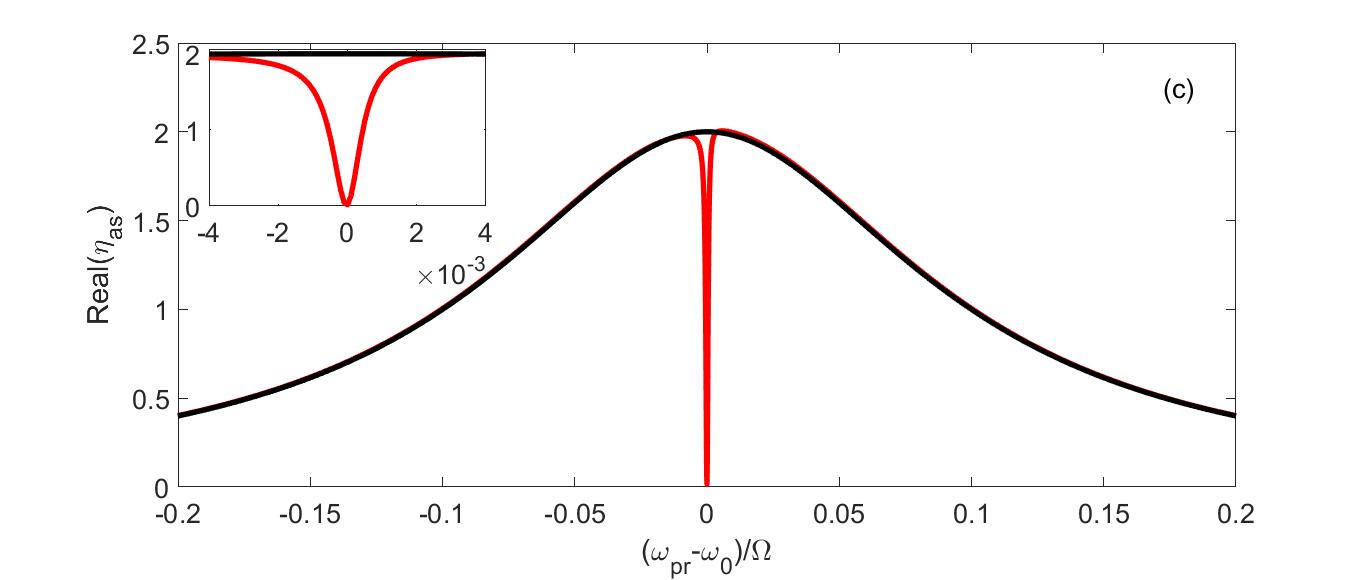}
	
	\caption{Real part of anti-Stokes field as a function of normalized probe detuning for $\omega_{m1}=\Omega$, $\phi_{m1}$=$3\pi/2$, (a) $s_{m1}$=$11$fN (blue), $6$fN (red), $0$ (black) and $s_{m2}$=$0$; (b) $s_{m1}$=$s_{m2}$=$5.5$ fN (blue), $s_{m1}$=$5.5$fN and $s_{m2}$=$0$ (red), $s_{m1}$=$s_{m2}$=$0$ (black); (c) $s_{m1}$=$11$fN and $s_{m2}$=$0$ (red), $s_{m1}$=$s_{m2}$=$11$fN, $\phi_{m2}$=$\pi/2$ (black).}
	\label{fig:OMIT} 
\end{figure}

\begin{figure*}
	\begin{center}
		\includegraphics[width=\linewidth,keepaspectratio]{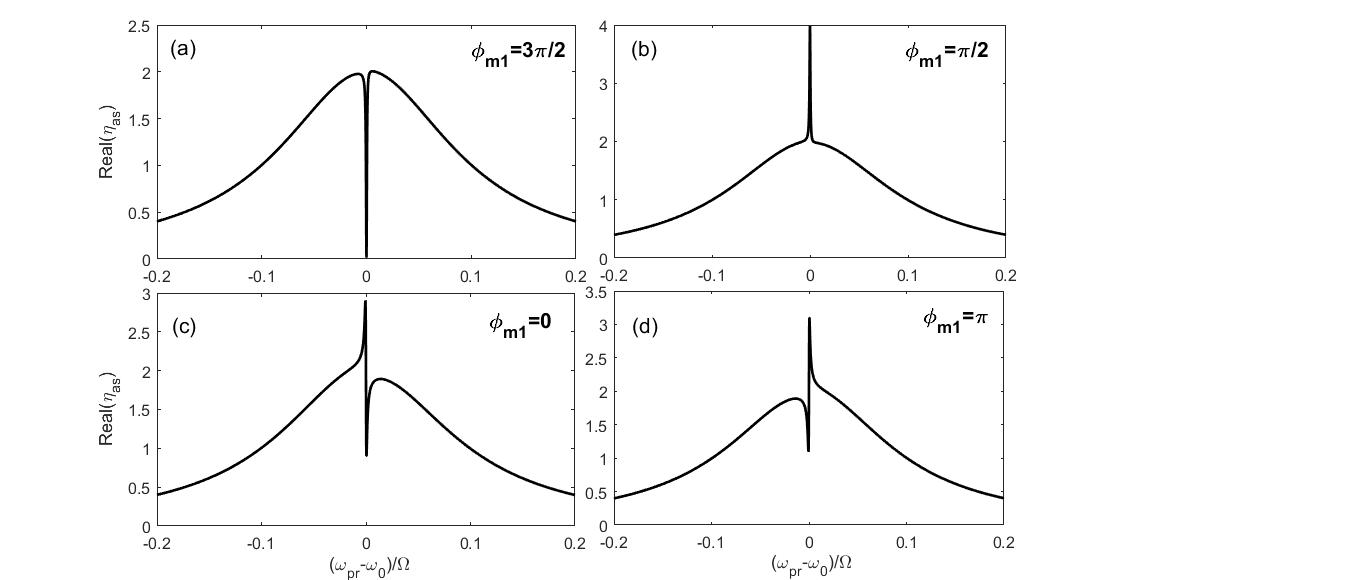}
				\caption{Real part of anti-Stokes field as a function of normalized probe detuning for different values of phase $\phi_{m1}$. $(a) \; \phi_{m1}=3\pi/2, (b) \; \phi_{m1}=\pi/2, (c) \; \phi_{m1}=0, (d) \; \phi_{m1}=\pi$
		}
		\label{fig:phase} 
	\end{center}
\end{figure*}

Taking  $\Delta=\delta=\Omega$,  it is observed that a phase of $3\pi/2$ gives rise to OMIT at the line center.  It is further  seen that the strength of OMIT feature (as quantified by how close the dip is to its zero value)  is proportional to the amplitude of the coherent mechanical pump which is illustrated  in Fig. \ref{fig:OMIT}(a).  Absence of mechanical driving i.e., $s_{m1}=0$ (black), shows absence of OMIT at the line center.   Introducing  a small driving amplitude of $s_{m1}=6$ fN (red)  results in  a dip at the line center, which indicates transmission of the probe beam on resonance.   With further increase in the amplitude, $s_{m1}=11$ fN (blue) one observes a complete transmission of probe beam.    The destructive interference between the probe beam and the anti-Stokes field when $\omega_{pr}=\omega_{pu}+\omega_{m1}$ gives rise to  OMIT at the line center corresponding to  $\omega_{m1}=\Omega$. The results of  Fig. \ref{fig:OMIT} show that an increase in the strength  of mechanical pump  gives rise to an enhancement in probe transmission, leading to complete transmission (OMIT) for a particular value of this parameter.  

In Fig. \ref{fig:OMIT}(b),  the absence of dip at the line center in the black curve shows clearly that OM effects are not present when the mechanical driving field is not applied, i.e., $s_{m1}=s_{m2}=0$.  Introduction of a small mechanical pump of amplitude  $s_{m1}=5.5$ fN on mirror 1, with $\omega_{m1}=\Omega$,  gives rise to a dip at the line center, as seen in the red curve, showing clearly the generation of OMIT.  Further inclusion of mechanical pump of the same amplitude ($s_{m2} = 5.5$ fN) on mirror 2, with both the phases held at the same value, i.e.,  $\phi_{m1} = \phi_{m2} = 3\pi/2$, gives rise to complete transparency of the probe beam  at the line center (blue).  This feature can be attributed as arising due to coherent addition of the OM contributions arising from each of the mechanical driving fields, which are at the same phase.

This can further be substantiated by exploring whether the effect will cancel out by tuning one of the mechanical driving  fields completely out of phase with the second.  This is indeed the case, as shown in  Fig. \ref{fig:OMIT}(c),  where the red curve shows complete transparency of probe beam at the line center,  generated solely due to $s_{m1}$,  at a phase $\phi_{m1} = 3 \pi/2$.    Introducing the second mechanical drive $s_{m2}$ at a phase $\phi_{m2} = \pi/2$ provides another channel which  destructively interferes  to cancel out this OMIT effect  as seen from the black curve. Thus, tuning the amplitude and phase of the coherent mechanical pumps  gives a handle to control the generation of  OMIT in a macroscopic cavity. 

In the results presented so far, it was shown that OMIT is observed at the line center, when the phases of the mechanical driving  fields were held at $3\pi/2$.  It would be interesting to see the effect of varying  the phase of the mechanical pump on the OMIT features.  For this purpose, at first, we switch off one of the mechanical pumps, say $s_{m2}=0$ and keep the amplitude $s_{m1}$ fixed at a particular value of $11$ fN  and vary it's  phase $\phi_{m1}$ at intervals of $\pi/2$ and record  the changes in the output spectrum which are  illustrated in Fig. \ref{fig:phase}.

 Fig. \ref{fig:phase}(a) illustrates the case when only one of the mechanical driving fields is turned on ($s_{m1}=11$ fN and $s_{m2}=0$) at a phase $\phi_{m1}=3\pi/2$, with  $Re(\eta_{as})$ reaching its minimum value at the line center, identical to the red curve in Fig. \ref{fig:OMIT}(c).  Keeping all other parameters fixed, we now change the phase $\phi_{m1}$ of $s_{m1}$ to $\pi/2$, which gives rise to a sharp increase in $Re(\eta_{as})$ at the line center as shown in  Fig. \ref{fig:phase}(b) showing remarkable absorptive behaviour of the cavity at $\pi/2$ phase. Other possible interesting values of phase  $\phi_{m1}$ are explored further.  The phase $\phi_{m1}=0$ results in a  Fano-like lineshape as shown in Fig. \ref{fig:phase}(c) and by changing the value of the phase  to $\phi_{m1}=\pi$, gives rise to the lineshape as shown in Fig. \ref{fig:phase}(d),  which is a mirror image of the previous case.  Similar features have been observed in other optomechanical systems  \cite{jma2015,jia2015}.  These results clearly show the importance of coherent mechanical pump and its phase in controlling the spectral features of the generated fields.  The sensitive changes in the behaviour of the system, which are detected as a function of the phase of mechanical driving, suggest that this method may be employed as a tool to detect the phase of an unknown harmonic force with considerable precision. 
 
\begin{figure}
	\includegraphics[width=\linewidth,keepaspectratio]{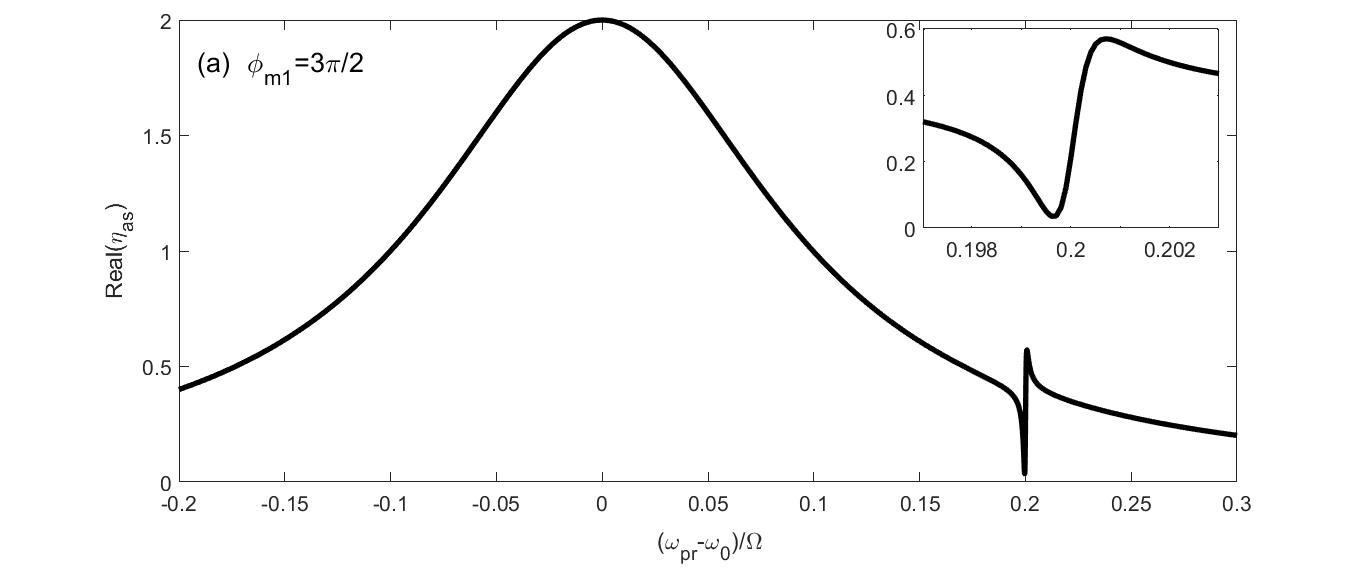}
	\includegraphics[width=\linewidth,keepaspectratio]{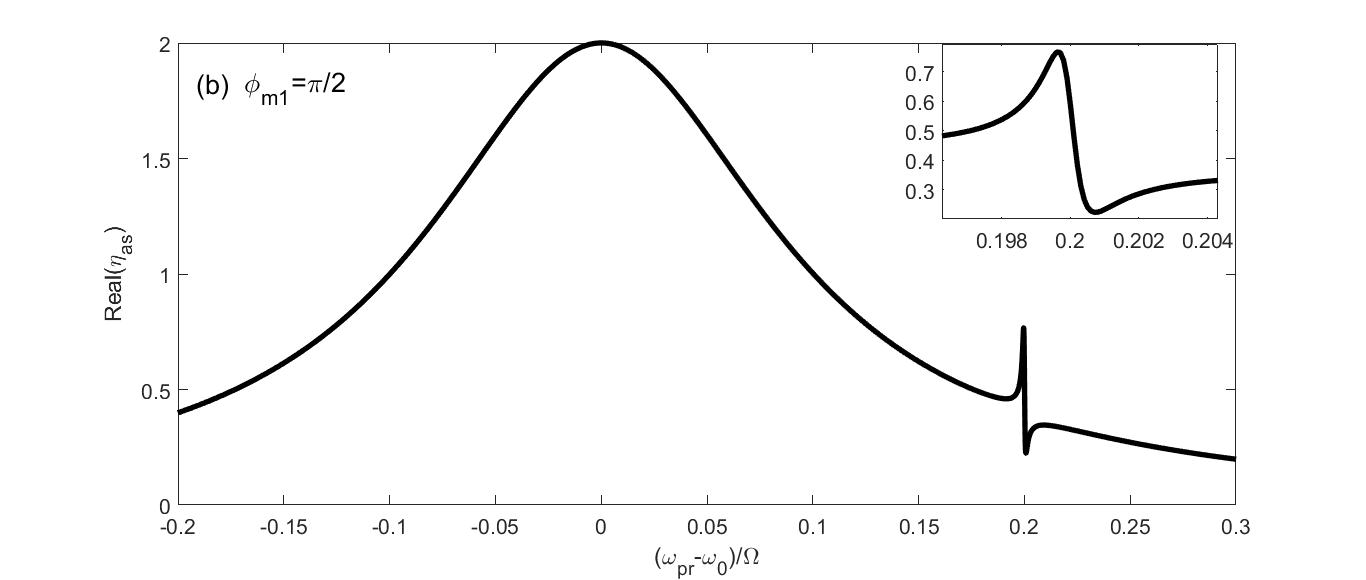}
	
	\caption{ Fano resonance in the real part of anti-Stokes field as a function of normalized probe detuning for  $s_{m1}=8$ fN and (a) $\phi_{m1}=3\pi/2$; (b) $\phi_{m1}=\pi/2$. 
	}
	\label{fig:phase_2} 
\end{figure}

In the above, we have assumed that the oscillation frequencies of both the mirrors are equal to the effective cavity detuning, i.e. $\omega_{m1} =\omega_{m2}= \Omega$, the combination of which was giving rise to resonance at the line center.  However, the ability to tune  the oscillation frequencies of movable mirrors independently of one  another can give rise to asymmetric lineshapes as shown in the  Figures \ref{fig:phase_2}(a) and (b).  Here we have tuned the oscillation frequency $\omega_{m1}=1.2 \Omega$, due to which the anti-Stokes field will destructively interfere with the probe beam  whenever $\omega_{pr} = \omega_{pu}+ 1.2 \Omega$,   as shown in Fig. \ref{fig:phase_2}(a). We observe occurrence of Fano-like resonance and a mirror image of the same, at a corresponding value of 0.2 of the normalized probe detuning,  for phases $\phi_{m1}=3\pi/2$  and $\pi/2$ respectively.  The results presented for various combinations of the mechanical drive fields and their phases show clearly the interference effects between two transition pathways,  in this case the fields generated at the probe frequency and at the anti-Stokes frequency.   It is interesting to note that these features have not been observed so far  in macroscopic cavities (due to the very weak optomechanical effects in such systems) and are resulting purely due to the introduction of coherent mechanical driving field(s).

\begin{figure}
	\includegraphics[width=\linewidth,keepaspectratio]{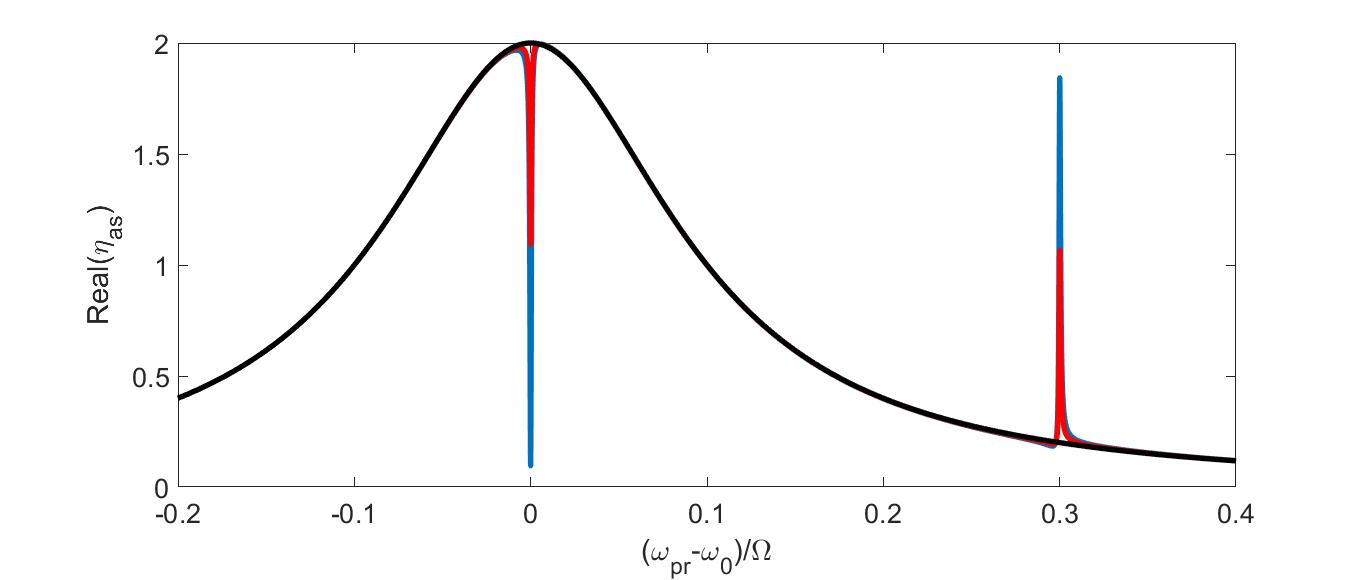}
	\caption{Real part of anti-Stokes field as a function of normalized probe detuning $(\omega_{pr}-\omega_{0})/\Omega$ for $s_{m1}=10.5$ fN, $\phi_{m1}=3\pi/2$  and $s_{m2}=38$ fN, $\phi_{m2}=\pi$ (blue), $s_{m1}=5$ fN, $\phi_{m1}=3\pi/2$ and $s_{m2}=20$ fN, $\phi_{m2}=\pi$ (Red); $s_{m1}=0$, $\phi_{m1}=3\pi/2$ and $s_{m2}=0$, $\phi_{m2}=\pi$ (black).
	}
	\label{fig:6} 
\end{figure}

Next, we consider the case when both the mechanical pumps ($s_{m1}$ and $s_{m2}$) are switched on and degeneracy between oscillation frequencies of the mirrors  is removed ($\omega_{m1} \ne \omega_{m2}$). This condition provides two distinct transition pathways (($\omega_{pu}+\omega_{m1})$ and $(\omega_{pu}+\omega_{m2}$)) to interfere with $\omega_{pr}$, which gives rise to two resonances as clearly illustrated in Fig. \ref{fig:6}.  Here, in addition to the OMIT generated at the line center, a sharp resonance peak at $\delta_{pr}/\Omega=0.3$ is observed. The parameter considered here are $\omega_{m1}= \Omega$, $\phi_{m1}= 3\pi/2$ and $\omega_{m2}=\Omega+0.3 \Omega$, $\phi_{m2}=\pi$.  Fig. \ref{fig:6},  clearly shows that the strength (magnitude) of these OM features can be controlled by tuning the amplitude of the mechanical driving.  When $s_{m1}=10.5$ fN, we see a minima in the  generated anti-Stokes field ($Re(\eta_{as})$) at the line center and a maxima at $\delta_{pr}/\Omega$= 0.3 for $s_{m2}=38$ fN, as shown by the blue curve in Fig. \ref{fig:6}. With a decrease in the amplitudes of both mechanical pumps, $s_{m1}=5$ fN and $s_{m2}=20$ fN,  we see a considerable decrease in the strength of resonant curves, as can be seen from the red curve.  The black curve illustrates the absence of any OM features as both the mechanical pumps are switched off, i.e., $s_{m1}=s_{m2}=0$.  It is to be mentioned that these OM features are generated in a macroscopic OM system with the introduction of coherent mechanical pump which has hitherto not been seen.

\begin{figure}
	\includegraphics[width=\linewidth,keepaspectratio]{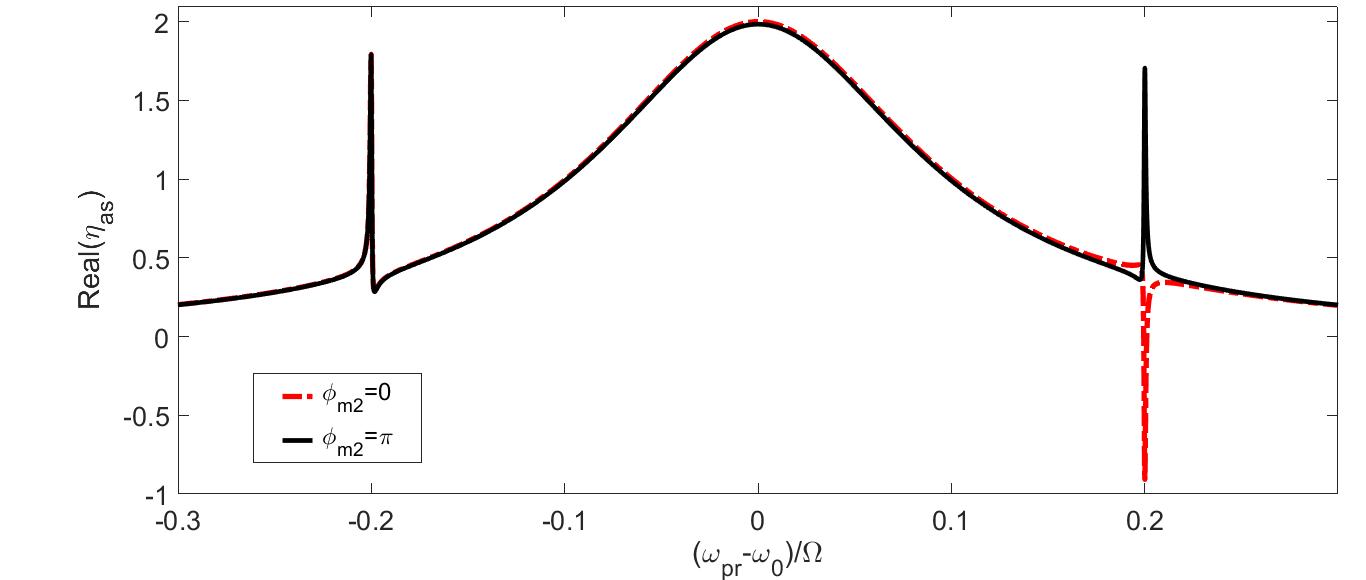}
	\caption{Double Fano resonance in real part of antistokes field as a function of normalized probe detuning for
		$s_{m1}=15$ fN, $\omega_{m1}=0.8\Omega$, $s_{m2}=20$ fN, $\omega_{m2}=1.2\Omega$ and $\phi_{m1}=\phi_{m2}=0$ (black solid curve); $\phi_{m1}=0$ and $\phi_{m2}=\pi$ (red dashed curve)
	}
	\label{fig:triple} 
\end{figure}

We next present our results on double Fano-like resonance lineshapes away from the line center, which can be tuned by introduction of mechanical driving fields and their phases.  
Such features have been widely studied  in a different context in plasmonic structures \cite{wang2013,dana2016}.   Very recently,  these double Fano resonance lineshapes have been studied also in cavity optomechanical systems \cite{qu2013,jiang2017,sohail2018}. In Fig. \ref{fig:triple} the resonant peak/dip on the right to  the line center  appears  due to the introduction of $s_{m1}$ and the peak on the  leftside  results due to the introduction of $s_{m2}$.  The parameters that were considered here are $\omega_{m1}=\Omega+0.2\Omega$ and $\omega_{m2}=\Omega-0.2\Omega$, due to which the resonance due to mirror 1 ($s_{m1}$) occurs at $\delta_{pr}/\Omega = 0.2$ and that due to mirror 2 ($s_{m2}$)  occurs at $\delta_{pr}/\Omega = -0.2$.
These resonances occur due to interference of probe beam and the anti-Stokes fields that are generated at $\omega_{pu}+\omega_{m1}$ and $\omega_{pu}+\omega_{m2}$ respectively.  The location of these peaks  can be suitably modified by tuning the mechanical frequency of the movable mirrors.  Here, a mechanical pump $s_{m1}=15$ fN (Fig. \ref{fig:triple}) is applied  to generate a strong resonance peak on the left, whereas a slightly larger value of  $s_{m2}$ ($20$ fN) is required  to generate a resonance peak of similar height on the right side.  This asymmetry arises from the fact that the resonance due to $s_{m2}$ occurs when $\omega_{pr}=\omega_{pu}+1.2\Omega$,  which is far away from the pump laser frequency as compared to the resonance that occurs due to application of $s_{m1}$ at $\omega_{pr}=\omega_{pu}+0.8\Omega$.  Therefore the closer we are to the pump laser frequency $\omega_{pu}$, the smaller is the force that is needed to generate the OM resonance features and vice versa.  One observes  that by flipping  the phase $\phi_{m2}$  from a value of $0$ (red dashed curve) to $\pi$ (black solid curve),  the dip changes to a peak. The negative peak (red dashed) at $\delta_{pr}/\Omega=0.2$ corresponds to resonance amplification which can effectively be shifted to resonance absorption (positive peak) by adjusting the phase ($\phi_{m2}$) of coherent mechanical pump.   These features clearly show the important role played by the amplitude and phase of the coherent mechanical pump in tuning the absorption/amplification features as well as their positions.

\begin{figure}
	\includegraphics[width=\linewidth,keepaspectratio]{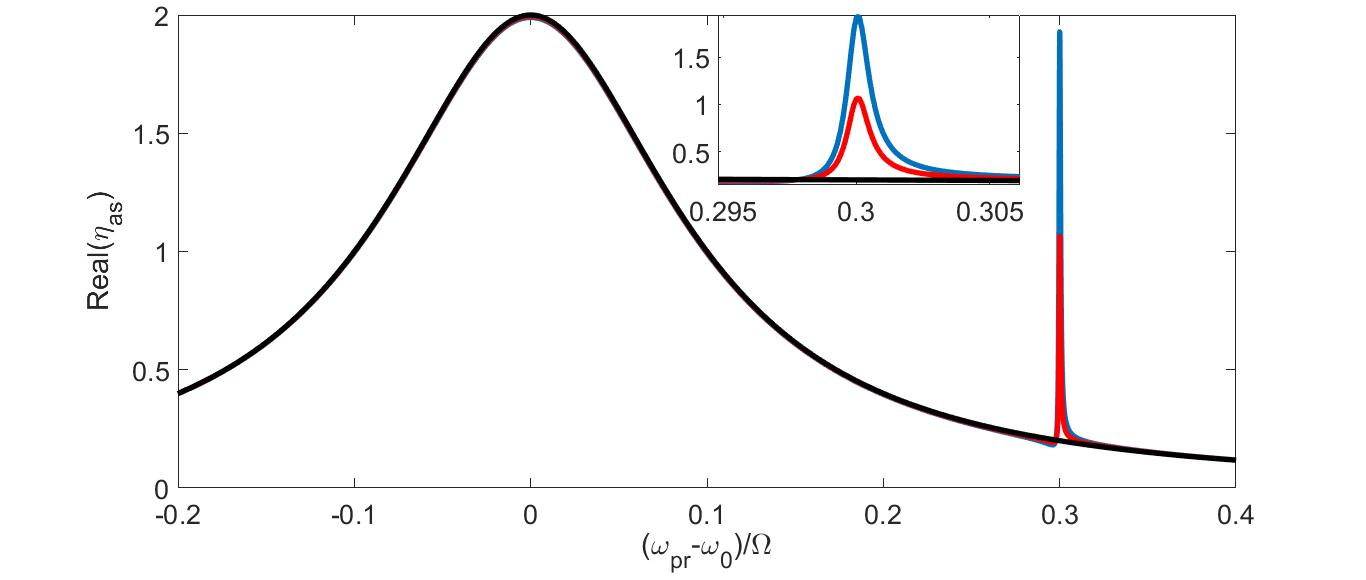}
	\caption{Real part of anti-Stokes field vs normalized probe detuning for $s_{m1} = s_{m2}=20$ fN, $\phi_{m1} =  \phi_{m2} = \pi$ (blue);  $s_{m1}=30$ fN, $\phi_{m1}= \pi$, $s_{m2}=10$ fN, $\phi_{m2}= 0$ (red);  $s_{m1} = s_{m2}=20$ fN, $\phi_{m1}= \pi$, $\phi_{m2}= 0$ (black) }
	\label{fig:fanophase} 
\end{figure}

 Next we consider the case when the mechanical frequency of both oscillators are taken to be equal and tuned away from the effective cavity detuning ($\Delta$) where $\omega_{m1}=\omega_{m2} = 1.3\Omega$ where $\Delta=\Omega$, this enables one to control the  resonance features by changing the phase of the mechanical pump. When both $s_{m1}$ and $s_{m2}$ have the same phase, namely,  $\phi_{m1}=\phi_{m2}=\pi$, they constructively interfere giving rise to resonant enhancement  at $\delta_{pr}/\Omega=0.3$ as shown in  the blue curve in Fig. \ref{fig:fanophase}. Next,  when the relative phase of the mechanical pumps, with strengths $s_{m1}=s_{m2}=20$ fN, is shifted from $\pi$ to $2\pi$ or $0$, destructive interference takes place between the two coherent processes which leads to total cancellation of the Fano-like feature, resulting in the black curve.  This situation amounts to effectively turning both the mechanical pumps off.  We thus show that by tuning  the  relative phase between the two mechanical pumps, which are of equal magnitude, we can completely switch on/off the Fano-resonance. Taking their amplitudes unequal will result in further features in the Fano-resonance.  For example,  the red curve in Fig. \ref{fig:fanophase} corresponds to unequal values of the mechanical driving fields,  $s_{m1}=30$ fN and $s_{m2}=10$ fN with phases $\phi_{m1}=\pi$ and $\phi_{m2}=0$, unlike the blue curve,  in which both the mechanical drives were  taken to have the same amplitude and phase.  
 
\section{Conclusion}
In this work we have shown how the OM features viz., the optomechanically induced transparency and asymmetric Fano lineshapes can arise in a four mirror macroscopic optomechanical cavity,  due to inclusion of  coherent mechanical driving of the two movable mirrors.  We identify interfering pathways leading to the Fano resonances, in the macroscopic four mirror optomechanical system considered here.  We further show that these features can be efficiently  controlled by changing the phase and amplitude of mechanical driving.   The sensitive changes that are observed in the Fano lineshapes with slight modification in amplitude and phase of the mechanical driving fields suggests the possibility of exploiting this feature to detect unknown harmonic forces.  For the special case of the frequencies of both the mechanical oscillators being equal, it is shown that the phase can be used as a switch to generate interesting optomechanical effects.  The freedom of tuning the two mechanical oscillators independently of each other,  leads to the generation of tunable double Fano-like resonance.  In conclusion, this work suggests the possibility of observing interesting tunable quantum effects at macroscopic scales, with the aid of coherent mechanical driving fields.

\end{document}